\documentclass[conference]{IEEEtran}
\usepackage{amsfonts}
\usepackage{amssymb}
\usepackage{cite}
\usepackage[cmex10]{amsmath}
\usepackage{float}
\usepackage{color}
\usepackage{stfloats,fancyhdr}

\usepackage{algorithm}
\usepackage{algorithmic}
\usepackage{multirow}
\usepackage{changepage}
\usepackage[normalem]{ulem}
\IEEEoverridecommandlockouts

\ifCLASSINFOpdf
  \usepackage[pdftex]{graphicx}
  \DeclareGraphicsExtensions{.pdf,.jpeg,.png}
\else
  \usepackage[dvips]{graphicx}
  \DeclareGraphicsExtensions{.eps}
\fi

\usepackage{subfigure}

\usepackage{fancybox,dashbox}

\begin{document}

\title{\huge{Exploiting Energy Accumulation Against Co-channel Interference in Wireless Energy Harvesting MIMO Relaying}
}

\author{Yifan Gu, He Chen, Yonghui Li, and Branka Vucetic
\\
School of Electrical and Information Engineering, The University of Sydney, Sydney, NSW 2006, Australia\\
Email: \{yifan.gu, he.chen, yonghui.li, branka.vucetic\}@sydney.edu.au
}

\maketitle

\begin{abstract}
This paper investigates a three-node multiple-input multiple-output relay system suffering from co-channel interference (CCI) at the multi-antenna relay. Contrary to the conventional relay networks, we consider the scenario that the relay is an energy harvesting (EH) node and has no embedded energy supply. But it is equipped with a rechargeable battery such that it can harvest and accumulate the harvested energy from RF signals sent by the source and co-channel interferers to support its operation. Leveraging the inherent feature of the considered system, we develop a novel accumulate-then-forward (ATF) protocol to eliminate the harmful effect of CCI. In the proposed ATF scheme, at the beginning of each transmission block, the relay can choose either EH operation to harvest energy from source and CCI or information decoding (ID) operation to decode and forward source's information while suffering from CCI. Specifically, ID operation is activated only when the accumulated energy at the relay can support an outage-free transmission in the second hop. Otherwise, EH operation is invoked at the relay to harvest and accumulate energy. By modeling the finite-capacity battery of relay as a finite-state Markov Chain (MC), we derive a closed-form expression for the system throughput of the proposed ATF scheme over mixed Nakagami-m and Rayleigh fading channels. Numerical results validate our theoretical analysis, and show that the proposed ATF scheme with energy accumulation significantly outperforms the existing one without energy accumulation.
\end{abstract}


\IEEEpeerreviewmaketitle

\section{Introduction}
Recently, wireless energy harvesting (WEH) technique has been regarded as a new viable solution to prolong the lifetime of energy constrained wireless networks. With this technique, wireless devices are enabled to extract the energy carried by radio-frequency (RF) signals to charge their batteries. In \cite{Nasir_time_switching}, Nasir \emph{et. al} first studied a classic three-node WEH relay network, wherein a WEH relay is deployed to assist the information transmission between a source-destination pair. Specifically, the relay is assumed to have no embedded energy supply and purely relies on the energy harvested from the RF signals broadcast by the source to support its operation. A practical receiver architecture, named time-switching (TS) relaying, was proposed in \cite{Nasir_time_switching} for the case that the relay is equipped with separate energy receiver and information receiver. In TS relaying, the received RF signals is either delivered to energy receiver for energy harvesting (EH) or be delivered to information receiver for information forwarding. Very recently, this seminal work has been extended to more general scenarios, see e.g., relay interference channel \cite{Chen_TWC_2015_Dis} and multi-relay networks \cite{YIGU4}.

In wireless networks, limited spectrum resources are normally reused to pursue higher spectrum efficiency. As a result, it is generally unavoidable for the desired links to suffer from co-channel interference (CCI). The performance degradation caused by CCI in conventional communication networks has been widely studied in the existing literature, see \cite{traditional_CCI} and references therein. However, only limited papers have investigated the impact of CCI on the performance of WEH networks. In \cite{CCI1}, the authors considered a single-input-single-output communication system with a WEH destination and proposed an opportunistic EH scheme to exploit the CCI as a potential energy source for the WEH device. Y. Chen also investigated a point-to-point communication network with CCI and revealed that the ratio of interference power to the source power is crucial to determine whether the CCI is beneficial or harmful to the system \cite{CCI3}. Very recently, \cite{CCI2} extended the work in \cite{CCI1} to a multiple-antenna relay network and three linear processing schemes, named maximum ratio combining (MRC), zero-forcing (ZF) and minimum mean-square error (MMSE) were investigated for the WEH relay, respectively. Results showed that implementing multiple antennas at the WEH relay can increase the EH capability and system performance significantly.

In \cite{CCI2}, it is assumed that the WEH relay exhausts the harvested energy during each transmission block to perform information transmission/forwarding. However, this assumption could lead to suboptimal system performance by realizing the following fact: when the CCI of the current transmission block is strong such that the relay can hardly decode the information from the source, the relay should harvest energy during the whole block and store the harvested energy for future information transmission instead of forwarding the source's information in the current block. At the same time, the relay can harvest substantial amount of energy from the CCI since the CCI is strong and this can improve the system performance in a long run. In this sense, the consideration and modeling of the energy accumulation (EA) process at the WEH relay is essential such that the relay is able to accumulate the harvested energy and use it efficiently to perform information transmission in an appropriate transmission block. To the best knowledge of the authors, there is no paper in open literature that studies the system performance of a WEH relay system with CCI and energy accumulation.

Motivated by the aforementioned gap, in this paper we focus on the design and analysis of a WEH MIMO relay system that exploits its energy accumulation capability to perform against the CCI. The main contributions of this paper are summarized as follows: {\textbf{(1)}} We develop an accumulate-then-forward (ATF) protocol for the considered WEH relay system to eliminate the effect of CCI. {\textbf{(2)}} By modeling the relay battery by a finite-state Markov Chain, we evaluate the transition matrix and stationary distribution of the relay battery. {\textbf{(3)}} Considering a decode-and-forward protocol implemented at the relay, we derive a closed-form expression for the system throughput over mixed Nakagami-m and Rayleigh fading channels. All theoretical analysis is then validated by numerical results. {\textbf{(4)}} Our results revealed that different from the conventional relay networks where CCI always degrades the system performance, the CCI may benefit the system performance in the considered WEH relay system. It is shown that the proposed ATF protocol with energy accumulation significantly outperforms the existing one without energy accumulation.

\textbf{\emph{Notation}}: Throughout this paper, we use $f_{X}(x)$ and $F_{X}(x)$ to denote the probability density function (PDF) and cumulative distribution function (CDF) of a random variable $X$. $\mathbb{E}\left\{\cdot \right\} $ represents the expectation operator. $\Gamma \left( {\cdot} \right)$ is the Gamma function \cite[Eq. (8.310)]{Gradshteyn_book_2007}, $\gamma \left( {\cdot,\cdot} \right)$ is the lower incomplete gamma function \cite[Eq. (8.350.1)]{Gradshteyn_book_2007} and $\Gamma \left( {\cdot},{\cdot} \right)$ is the upper incomplete gamma function \cite[Eq. (8.350.2)]{Gradshteyn_book_2007}. $\left( \cdot \right)^T$ and $\left( \cdot \right)^\dag$ represent the transpose and conjugate transpose, respectively. ${{\left\| \cdot \right\|}}$ is the Frobenius norm and ${\bf I}_{N}$ denotes the $N \times N$ identity matrix.

\section{System Model and Protocol Description}
\subsection{System Description}
In this paper, we investigate a dual-hop MIMO relay system consists of one single-antenna source $S$, one single-antenna destination $D$ and one decode-and-forward (DF) relay $R$ equipped with $N$ antennas. We assume that there is no direct link between $S$ and $D$ due to obstacles or severe attenuation, and thus the communication can be established only via the relay. Unlike the conventional relay networks, in this paper we consider that $R$ have no embedded power supply and it only relies on the energy harvested from the RF signals broadcast by $S$ before helping forward its signal to $D$. Besides, An energy storage is also assumed to be equipped at $R$ such that it can perform energy accumulation and scheduling across different transmission blocks. As in \cite{CCI2}, we assume that the relay is suffered from $L$ co-channel interferers (denoted by ${I_1}, \ldots ,{I_L}$) and additive white Guassian noise (AWGN) while the destination is still corrupted by the AWGN only\footnote{The scenario where relay and destination experience different interference will occur in frequency-division relaying systems \cite{CCIONLY}.}.

\subsection{Channel Model}
Since the up-to-date wireless energy transfer techniques could only be operated within a relatively short communication range and the line-of-sight (LoS) path is very likely to exist, Rician fading would be the most appropriate one to characterize the channel fading of $S-R$ link in the first hop. However, the statistical functions (e.g., CDF and PDF) of Rician fading are very complicated, which would make the ensuing analysis extremely difficult \cite{Zhong_Tcom_2015_Wireless}. Fortunately, the Rician distribution could be well approximated by the more tractable Nakagami-m fading model. Thus, in this paper we adopt the Nakagami-m distribution to model the channel fading for $S-R$ link. As the distance between $R-D$, $R-I_l$, $l \in \left\{1,2, \ldots,L\right\}$ can be much further, all the other links in the system are considered to be subject to Rayleigh fading.

Let ${\bf{h}}_0\in \mathbb{C}^{N \times 1}$ and ${\bf{g}}_0\in \mathbb{C}^{N \times 1}$ denote the complex channel vectors of $S$-$R$ and $R$-$D$ links, respectively. Moreover, we use ${\bf{h}}_l\in \mathbb{C}^{N \times 1}$, $l \in \left\{1,\ldots,L\right\}$, to represent the complex channel vector from interferer $I_l$ to $R$.  The magnitudes of all entries of ${\bf{h}}_0$ are assumed to follow an independent and identical distributed (i.i.d.) Nakagami-m distribution with a shape parameter\footnote{For the purpose of exploration, we consider $m$ to be integer in this paper.} $m$ and an average power gain $\Omega_0$. On the other hand, each element of ${\bf{g}}_0$ and ${\bf{h}}_l$ is subjected to an i.i.d. circularly symmetric complex Gaussian (CSCG) distribution with zero mean and variance $\Lambda_0$ and $\Omega_l$, respectively. We also assume that all desired and interfering links in the MIMO relaying system experience independent slow and frequency flat fading, where channel gains remain constant during each transmission block, denoted by $T$, but change independently from one block to another.

\subsection{Accumulate-Then-Forward Protocol}
At the beginning of each transmission block, the relay can choose either EH operation to accumulate energy or information decoding (ID) operation to forward source's information. Specifically, each transmission block is divided into two time slots with equal length $T/2$. If the relay's accumulated energy is sufficient to support an outage-free transmission in the second hop, it will choose the ID operation and decode the received signal from $S$ while suffering from CCI and AWGN in the first time slot. Otherwise, $R$ will perform EH operation to harvest energy from both the RF signals transmitted by $S$ and co-channel interferers. In the proposed ATF protocol, for the operation simplicity of the energy constrained relay, we restrict that $R$ provides no feedback to $S$ and $S$ always remains silent during the second time slot. As a result, $R$ will continue to solely harvest energy from CCI signals in the second time slot if EH mode is invoked. Moreover, if $R$ chooses ID mode but an incorrect decoding occurs, it will also harvest energy from CCI signals during the second time slot. On the other hand, when $R$ performs ID during the first time slot and the information is decoded correctly, $R$ will forward the source's information packet to $D$ by depleting the accumulated energy in its battery.


The received signal at $R$ during the first time slot can be characterized as
\begin{equation}\label{eq:received_signal_R}
{\bf{y}}_R = \sqrt {{P_0}}  {\bf{h}}_0 {x_0} + \sum\limits_{l = 1}^L {\sqrt {{P_l}} {{\bf{h}}_l}{x_l}}  + {{\bf{n}}_R},
\end{equation}
where $x_0$ and $x_l$ are the signals transmitted by $S$ and $I_l$ with unit energy, $P_0$ and $P_l$ are the transmit power of $S$ and $I_l$ respectively, and ${{\bf{n}}_R}$ is an $N \times 1$ vector denotes the AWGN suffered at the relay with $\mathbb{E} \left\{{{\bf{n}}_R}{{\bf{n}}_R}^\dag\right\}=\sigma_R^2{\bf I}_{N}$. 
If $R$ do not have enough energy to support an outage-free transmission, EH mode is invoked and it harvests energy from (\ref{eq:received_signal_R}) during the first time slot. In the second time slot, when $S$ keeps silent, the received signal at $R$ is given by
\begin{equation}\label{2}
{\bf{y}}_R =\sum\limits_{l = 1}^L {\sqrt {{P_l}} {{\bf{h}}_l}{x_l}}  + {{\bf{n}}_R}.
\end{equation}

The relay will continue to harvest energy from (\ref{2}) during the second time slot. With out loss of generality, we consider a normalized transmission block (i.e., $T=1$) hereafter, the total amount of harvested energy for EH operation can thus be written as \cite{Louie_TWC_2009,Zhang_TWC_2013_MIMO}
\begin{equation}\label{eq:harvested_enery_ModeI}
   \tilde E_{\rm I} =  \frac{\eta}{2}{{P_0}{{\left\| {{{\bf{h}}_0}} \right\|}^2} }  + \eta \sum\limits_{l = 1}^L {{P_l}{{\left\| {{{\bf{h}}_l}} \right\|}^2}},
\end{equation}
where $0< \eta \le 1$ is the energy conversion efficiency of the energy receiver. In (\ref{eq:harvested_enery_ModeI}), we ignore the amount of energy harvested from the noise as it is normally below the sensitivity of EH circuit.

On the other hand, when $R$ has sufficient energy, it chooses ID operation to decode source's information. For a multi-antenna receiver corrupted by CCI, linear diversity combining schemes are normally adopted to alleviate the signal degradation \cite{Shah_TVT_2000_Per,Romero_TWC_2008_Rece}. In this paper, we assume that the maximum ratio combining (MRC) scheme is implemented at $R$ for its low complexity. Note that the adopted MRC scheme is particularly suitable for the considered system as it only requires the instantaneous CSI of $S-R$ link and it is difficult for the WEH relay to acquire the full CSI of all the interferers in the considered network.

In an MRC receiver, the antenna array elements are weighted by the channel gains and given by ${\bf w} = {\bf{h}}_0^\dag / {{\left\| {{\bf{h}}_0} \right\|}}$ \cite{Romero_TWC_2008_Rece}. Multiply the received signal given in (\ref{eq:received_signal_R}) by the antenna weights vector ${\bf w}$ and after some mathematical manipulation, the resultant SINR at $R$ is given by
\begin{equation}\label{eq:SINR_R}
\gamma _R = \frac{{{P_0}{{{\left\| {{\bf{h}}_0} \right\|}}^2}}}{{\sum\limits_{l = 1}^L {{P_l}{{\left\| {{{{v}}_l}} \right\|}^2}}+\sigma _R^2}},
\end{equation}
where ${v}_l = {\bf{h}}_0^\dag {{{\bf{h}}_l}} / {{\left\| {{{\bf{h}}_0}} \right\|}}$ and it is studied in \cite{v} that if the elements of ${{{\bf{h}}_l}}$ follow an i.i.d. CSCG distribution, then ${v}_l$ and ${{{\bf{h}}_0}}$ are mutually independent and ${{\left\| {{{{v}}_l}} \right\|}^2}$ is exponentially distributed with a variance $\Omega_l$.

If the decoding is incorrect, $R$ keeps harvest energy from the CCI during the second time slot and the amount of harvested energy can be characterized from (\ref{2}) as
\begin{equation}\label{eq:harvested_enery_ModeII}
\tilde E_{\rm II} = \frac{\eta }{2} \sum\limits_{l = 1}^L {{P_l}{{\left\| {{{\bf{h}}_l}} \right\|}^2}}.
\end{equation}

When the received information is decoded correctly, $R$ will forward the information to $D$ during the second time slot. Assume the maximum ratio transmission (MRT) scheme is implemented at the multiple antenna relay, the received signal-to-noise ratio (SNR) at $D$ is given by
\begin{equation}\label{eq:SINR_D_ZF}
\gamma _D = \frac{{{P_R}{{\left\| {{{\bf{g}}_0}} \right\|}^2}}}{{\sigma _D^2}},
\end{equation}
where $P_R$ is the transmit power of the relay and ${{\sigma _D^2}}$ is the variance of AWGN suffered at $D$.

To ensure an outage-free transmission, the optimal transmit power of the relay can be evaluated from the channel capacity and given by
\begin{equation}\label{eq:relay_power_threecase}
 P_R = \frac{{\mu \sigma _D^2}}{{\left\| {{{\bf{g}}_0}} \right\|}^2},\\
\end{equation}
where $\mu = 2^{2\mathbb R} -1$ and $\mathbb R$ is the transmission bit rate of $S$.

\section{Performance Analysis}
In this section, we first calculate the statistical distributions of the harvested energy ${\tilde E_{\rm I}}$ and ${\tilde E_{\rm II}}$ given in (\ref{eq:harvested_enery_ModeI}) and (\ref{eq:harvested_enery_ModeII}). We then derive the distribution of the optimal transmit power of the relay from the expression given in (\ref{eq:relay_power_threecase}). Subsequently, we characterize the dynamic charging/discharging behaviors of the relay battery by a finite state Markov chain (MC) and evaluate the transition probabilities of the formulated MC. With the aid of the MC model, we finally derive a closed-form expression for the system throughput of the proposed ATF protocol subject to CCI.

\subsection{Distributions}
Since each element of ${{{\bf{h}}_0}}$ is subject to an i.i.d. Nakagami-m distribution with a shape parameter $m$ and average power gain $\Omega_0$. The first term of the harvested energy in (\ref{eq:harvested_enery_ModeI}), i.e., $\frac{\eta}{2}{{P_0}{{\left\| {{{\bf{h}}_0}} \right\|}^2} }$, thus follows a gamma distribution with a shape parameter $mN$ and average power gain $\frac{\eta}{2}{{P_0}N\Omega_0}$. Each element in the second summation term $\eta {{P_l}{{\left\| {{{\bf{h}}_l}} \right\|}^2}}$ is a sum of $N$ i.i.d. exponential random variables and also follows a gamma distribution with a shape parameter $N$ and average power gain $\eta {{P_l}N\Omega_l}$. The harvested energy for EH operation $\tilde E_{\rm I}$ is a sum of $L+1$ mutually independent gamma random variables. To the best knowledge of authors, it is very complicated to characterize a closed-form expression for the exact distribution of $\tilde E_{\rm I}$. Fortunately, according to \cite{Costa_CL_2011_Int}, $\tilde E_{\rm I}$ can be tightly approximated as a single gamma random variable with the CDF given by
\begin{equation}\label{eq:CDF_EI}
\begin{split}
 {F_{\tilde E_{\rm I}}}\left( x \right)  \approx {{\gamma \left( {{m_I},{{{m_I}} \over {{\Omega _I}}}x} \right)} \over {\Gamma \left( {{m_I}} \right)}},
  \end{split}
\end{equation}
where parameters $m_I$ and $\Omega_I$ can be calculated by moment-based estimators \cite{Costa_CL_2011_Int}. Specifically, we have
\begin{subequations}\label{}
\begin{align}
&{\Omega _I} = \mathbb{E}\left[ {{{\tilde E}_{\rm{I}}}} \right],\\
&{m_I} = \frac{{\Omega _I^2}}{{\mathbb{E}\left[ {{{\tilde E}_{\rm{I}}}^2} \right] - \Omega _I^2}}.
\end{align}
\end{subequations}
Let $m_j$ and $a_j$, $j=1,2,\cdots,L+1$ denote the shape parameters and average power gains of the considered $L+1$ gamma random variables described above, respectively. It is easily attain that ${\Omega _I} = \sum\limits_{j = 1}^{L+1} {a_j} $. To obtain the value of  $m_I$, a multinomial expansion is implemented to express the moment term $\mathbb{E}\left[ {{{\tilde E}_{\rm{I}}}^n} \right]$ in terms of the individual moments of the summands given by
\begin{equation}
{m_I} = {{{\Omega _I}^2} \over {\Phi  + {\Omega _I}^2}}.
\end{equation}
The parameter $\Phi$ in the above expression is given by
\begin{equation}
\begin{split}
\mathbb{E}\left[ {{{\tilde E}_{\rm{I}}}^n} \right]  & = \sum\limits_{{n_1} = 0}^n { \sum\limits_{{n_2} = 0}^{{n_1}} {\cdots  \sum\limits_{{n_L} = 0}^{{n_{L - 1}}} {\binom{n}{n_1}\binom{n_1}{n_2} \cdots \binom{n_{L-1}}{n_{L}}} } }\\
& \quad \times \mathbb{E}\left[ {\beta _1^{n - {n_1}}} \right]\mathbb{E}\left[ {\beta _2^{{n_1} - {n_2}}} \right] \cdots \mathbb{E}\left[ {\beta _{L + 1}^{{n_L}}} \right],
\end{split}
\end{equation}
where
\begin{equation}
\mathbb{E}\left[\beta _j^n\right] = {{\Gamma \left( {{m_j} +  {n} } \right)} \over {\Gamma \left( {{m_j}} \right)}}{\left( {{{{a _j}} \over {{m_j}}}} \right)^{{n }}}.
\end{equation}

We now turn to derive the statistics of $\tilde E_{\rm II}$, which is a sum of $L$ gamma random variables with identical shape parameters $N$ and average power gains
 $\frac{\eta}{2}{{P_l}N\Omega_l}$. Similar to $\tilde E_{\rm I}$, the CDF of $\tilde E_{\rm II}$ can be summarized as
\begin{equation}\label{}
\begin{split}
 {F_{\tilde E_{\rm II}}}\left( x \right)  \approx {{\gamma \left( {{m_J},{{{m_J}} \over {{\Omega _J}}}x} \right)} \over {\Gamma \left( {{m_J}} \right)}},
  \end{split}
\end{equation}
where parameters $m_J$ and $\Omega_J$ can be evaluated similarly as in (9).

We then present the distribution of the required transmit power ${P_R}$. It is easily obtain that ${{\left\| {{{\bf{g}}_0}} \right\|}^2}$ follows a gamma distribution with a shape parameter $N$ and average power gain $N\Lambda_0$. From (\ref{eq:relay_power_threecase}), the CDF of ${P_R}$ can thus be expressed as
\begin{equation}
{F_{P_R}}\left( x \right) = \sum\limits_{i = 0}^{N - 1} {{{{{\left( {{{\mu \sigma _D^2}} \over {{\Lambda_0x}}} \right)}^i}} \over {i!}}} \exp \left( { - {{{\mu \sigma _D^2}} \over {{\Lambda_0x}}}} \right).
\end{equation}

\subsection{Markov Model of Relay battery}
To analyze the performance of the considered network with the ATF protocol, we subsequently model the dynamic charging/discharging behavior of the relay battery by a finite-state MC. In this paper, we follow \cite{Krikidis_CL_2012_RF} and adopt a discrete-level battery model. More specifically, We use $C$ to denote the capacity of the relay battery and $Q$ to denote the number of discrete energy levels excluding the empty level. Then, the $i$th energy level of the relay battery can be expressed as ${\varepsilon_i} = iC/Q$, $i  \in \left\{  0,1,2 \cdots Q\right\}$. We define state ${S_i}$ as the relay residual energy in the battery being ${\varepsilon_i}$. The transition probability $T_{i,j}$ is defined as the probability of the transition from state $S_i$ to state $S_j$. With the adopted discrete-level battery model, the amount of harvested energy ${\tilde E_{\rm I}}$ and ${\tilde E_{\rm II}}$ should be discretized to one specific energy level of the battery. The discretized amount of harvested energy is defined as
\begin{equation}\label{eq:discrete_EI}
{E_{\rm I}} \buildrel \Delta \over = {\varepsilon _{j}}, \quad {j} = \arg \mathop {\max }\limits_{i \in \left\{ {0,1, \cdots ,Q} \right\}} \bigg\{{\varepsilon _i}:{\varepsilon _i} <  {\tilde E_{\rm I}}\bigg\},
\end{equation}
\begin{equation}\label{}
{ E_{\rm II}} \buildrel \Delta \over = {\varepsilon _{j}}, \quad {j} = \arg \mathop {\max }\limits_{i \in \left\{ {0,1, \cdots ,Q} \right\}} \bigg\{{\varepsilon _i}:{\varepsilon _i} <  {\tilde E_{\rm II}}\bigg\}.
\end{equation}

On the other hand, when $D$ adopts ID operation and decode the received information correctly, the discretized value of the required outage-free energy at the relay is defined as
\begin{equation}\label{}
{E_R} \buildrel \Delta \over =
\left\{
\begin{matrix}
\begin{split}
&{\varepsilon _{j}}, \quad \text{if} \quad {{P_R} \over 2} \le C \\
& \infty, \quad \text{if} \quad {{P_R} \over 2} > C
\end{split}
\end{matrix}
\right.,
\end{equation}
where $j= \arg \mathop {\min }\limits_{i \in \left\{ {0,1, \cdots ,Q} \right\}} \bigg\{{\varepsilon _i}:{\varepsilon _i} \ge {{P_R}/2}\bigg\}$, and $P_R$ is given in (\ref{eq:relay_power_threecase}). Note that the factor $\frac{1}{2}$ is introduced because the relay performs information forwarding only during the second time slot.

With the operation principle of the proposed ATF scheme described in Sec. II and the MC model defined above, we can now ready to evaluate the state transition probabilities of the MC for the relay battery. Inspired by the seminal works given in \cite{Krikidis_CL_2012_RF}, all the transition probabilities of the considered model can be summarized into seven general cases given in (\ref{eq:Tij}) on top of next page: (\text{1}) The empty battery is partially charged. (\text{2}) The empty battery is fully charged. (\text{3}) The non-full and non-empty battery remains unchanged. (\text{4}) The battery remains full. (\text{5}) The non-empty battery is partially charged. (\text{6}) The non-empty battery is fully charged. (\text{7}) The battery is discharged.

In the following, we analyze case (5), which is the most complicated case. The non-empty battery is partially charged happens from two possible events. The first one is that the stored energy at the relay can not support an outage free forwarding operation, and EH operation is thus invoked and the relay harvests energy from both $S$ and co-channel interferers to charge its battery. The second event is that the relay has sufficient energy for an outage-free forwarding operation but it cannot decode the received information correctly. The relay thus harvests energy from only the $CCI$ during the second time slot to charge the battery. With the distributions evaluated in Sec. III-A, the transition probabilities for this case can be characterized in the 5th line of (\ref{eq:Tij}), where the terms $\left[1-{F_{P_R}}\left( 2iC \over Q \right)\right]$ and ${F_{P_R}}\left( 2iC \over Q \right)$ represents the probability that the stored energy at the relay $iC/Q$ cannot or can support an outage-free transmission, respectively. The terms $\left[{F_{\tilde E_{\rm I}}}\left( (j-i+1)C \over Q \right)- {F_{\tilde E_{\rm I}}}\left( (j-i)C \over Q \right)\right]$ and $\left[{{ F}_{\tilde E_{\rm II}}}\left( (j-i+1)C \over Q \right)-{{ F}_{\tilde E_{\rm II}}}\left( (j-i)C \over Q \right)\right]$ indicate that the harvested energy is discretized to $\left(j-i\right)$ energy levels of the relay battery so that the battery is charged from energy level $i$ to $j$. The term $\Pr \left\{ \rm O \right\}$ is the outage probability for the first hop with the received SINR given in (\ref{eq:SINR_R}) and it has been widely studied in the existing literature for conventional dual-hop relaying networks suffering from CCI. A closed-form expression of $\Pr \left\{ \rm O \right\}$ can be found in  \cite[Eq. (21)]{Romero_TWC_2008_Rece} and we omit it here due to space limitation.

For the other six transition cases, we omit the details of the derivation for transition probabilities and they can be evaluated similarly from the analysis given above for case 5.

\begin{figure*}[!t]
\begin{equation}\label{eq:Tij}
\begin{split}
{T_{i,j}}=\left\{ {
 \begin{matrix}
 \begin{split}
   &{{F_{\tilde E_{\rm I}}}\left( (j+1)C \over Q \right) - {F_{\tilde E_{\rm I}}}\left( jC \over Q \right)} , i=0, 0 \le j<Q \\
   &{1-{F_{\tilde E_{\rm I}}}\left( C \right)},i=0,j=Q  \\
   &{\left[1-{F_{P_R}}\left( 2iC \over Q \right)\right]{F_{\tilde E_{\rm I}}}\left( C \over Q \right)+\left[{F_{P_R}}\left( 2iC \over Q \right)\right]{{ F}_{\tilde E_{\rm II}}}\left( C \over Q \right)\Pr \left\{ \kappa \left[{n}\right] =0 \right\}}, 0<i=j<Q \\
   &{\left[1-{F_{P_R}}\left( 2C   \right)\right]+{F_{P_R}}\left( 2C  \right)\Pr \left\{ \rm O \right\}}, i=j=Q \\
   &{\left[1-{F_{P_R}}\left( 2iC \over Q \right)\right]\left[{F_{\tilde E_{\rm I}}}\left( (j-i+1)C \over Q \right)- {F_{\tilde E_{\rm I}}}\left( (j-i)C \over Q \right)\right] +}\\
   &{\Pr \left\{ \rm O \right\} {F_{P_R}}\left( 2iC \over Q \right)\left[{{ F}_{\tilde E_{\rm II}}}\left( (j-i+1)C \over Q \right)-{{ F}_{\tilde E_{\rm II}}}\left( (j-i)C \over Q \right)\right]}, 0<i<j<Q \\
   &{\left[1-{F_{P_R}}\left( 2iC \over Q \right)\right]\left[1- {F_{\tilde E_{\rm I}}}\left( (Q-i)C \over Q \right)\right] + \Pr \left\{ \rm O \right\} \left[{F_{P_R}}\left( 2iC \over Q \right)\right]\left[1-{{ F}_{\tilde E_{\rm II}}}\left( (Q-i)C \over Q  \right)\right]}, 0<i<Q, j=Q \\
   &{\left(1-\Pr \left\{ \rm O \right\}\right)\left[{F_{P_R}}\left( 2(i-j)C \over Q \right)-{F_{P_R}}\left( 2(i-j-1)C \over Q \right)\right] }, 0\le j<i \le Q\\
 \end{split}
\end{matrix}
  } \right..
\end{split}
\end{equation}
\hrulefill
\vspace*{4pt}
\end{figure*}

\subsection{System Throughput}
We now define $\mathbf {Z}= ({T_{i,j}})$ to denote the $(Q+1) \times (Q+1)$ state transition matrix of relay battery. By using similar methods in \cite{Krikidis_CL_2012_RF}, we can readily verify that the transition matrix $\mathbf{Z}$ derived from the above MC model is irreducible and row stochastic. Hence, there must exist a unique stationary distribution $\pmb{\pi}$ that satisfies the following equation
\begin{equation}\label{eq:steady_state}
\pmb {\pi}  =\left( {{{\pi} _{0}},{{\pi} _{1}}, \cdots, {{\pi} _{Q}}} \right)^{T} = \left({\mathbf{Z}}\right)^{T} \pmb {\pi},
\end{equation}
where ${{\pi} _{i}}$, $i \in \left\{ {0,1, \cdots ,Q} \right\}
$, is the $i$-th component of $\pmb {\pi}$ representing the stationary distribution of the $i$-th energy level at relay. The battery stationary distribution can be solved based on (\ref{eq:steady_state}) and expressed as 
\begin{equation}\label{app-energy-distri}
 \pmb {\pi} = {\left( { {\mathbf{Z}}^{T} - \mathbf{I} + \mathbf{B} }\right)^{ - 1}}\mathbf{b},
\end{equation}
where ${\mathbf{B}_{i,j}}=1, \forall i, j$ and $\mathbf{b}={(1,1, \cdots ,1)^T}$.

Since each discharge of battery guarantees an outage-free communication, the system outage probability of the proposed ATF scheme can be characterized as
\begin{equation}
P_{out} = \sum\limits_{i = 0}^Q {{\pi _i}\sum\limits_{j = i}^Q {{T_{i,j}}} }.
\end{equation}
We finally can obtain a closed-form expression of the system throughput of the proposed ATF protocol subject to CCI given by
\begin{equation}\label{throughput}
\Upsilon = \mathbb{R} \left(1- P_{out}\right).
\end{equation}

\section{Numerical Results}
In this section, we present some numerical results to validate and illustrate the above theoretical analysis. In order to capture the effect of path-loss, we model the desired channels as ${\Omega _{XY}} = {1 \over {1 + {d_{XY}^\alpha}}}$, where ${\Omega _{XY}}$ is the average channel power gain between node $X$ and $Y$, ${d_{XY}}$ denotes the distance between node $X$ and $Y$, and $\alpha  \in \left[ {2,5} \right]$ is the path-loss factor. For simplicity, we consider a linear topology such that the relay, source and destination are located in a straight line. Without loss of generality, the interferers are assumed to have identical transmit power denoted by $P_I$ but they are located differently and the distances between interferers and relay is denoted by the vector ${\bf{d}}_{IR}$. In all the following simulations, we set the distances ${d_{SD}} = 20$m,  ${d_{SR}} = 6$m,  ${\bf{d}}_{IR}=(12,13,14)$m, the shape parameter $m=2$, the path-loss factor $\alpha  = 2$, relay antenna number $N=4$, the battery capacity $C=0.5$, the battery levels $Q=90$, the noise powers ${\sigma_R^2} = \sigma_D^2= - 80$dBm and the energy conversion efficiency $\eta  = 0.5$.

\begin{figure}\label{fig:montesimu1}
\centering \scalebox{0.55}{\includegraphics{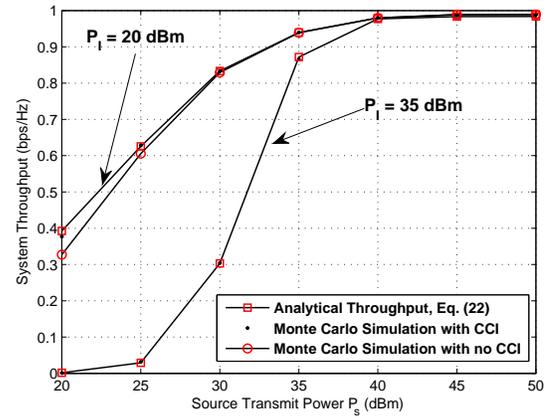}}
\caption{The system throughput of the proposed ATF scheme versus source transmit power $P_s$ for different interfererence power $P_I$. \label{fig:system}}
\end{figure}

We first compare the analytical system throughput derived in (\ref{throughput}) with the Monte Carlo simulation result. From Fig. 1, we can see that the analytical results coincide well with the Monte Carlo simulation which validates our theoretical analysis presented in Sec. II-III. Moreover, the system throughput becomes saturated when the transmit power is sufficiently large. This is due to the finite storage capacity of the relay battery in the sense that the accumulated energy cannot exceed the battery capacity. Furthermore, we can observe from this figure that when the interference power $P_I=20$dBm, the system throughput of the proposed ATF scheme is higher than the case without CCI. This is opposite to the conventional communication networks where the CCI always leads to performance degradation. However, when the interference power is increased to 40dBm, the performance of the proposed ATF scheme is obviously degraded. This observation can be explained as follows. The CCI signals are actually a `double-edged sword' in the considered system. On one hand, it deteriorates the quality of received signals at $R$ performing ID, which is harmful to the system performance. On the other hand, the CCI contributes extra amount of energy when $R$ harvests energy from the RF signals. This can enhance the energy accumulation speed at $R$, which is beneficial to the system performance.

\begin{figure}\label{fig:compare1}
\centering \scalebox{0.55}{\includegraphics{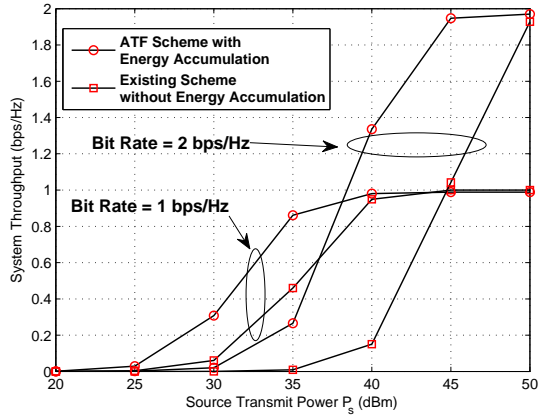}}
\caption{The system throughput of the proposed ATF scheme with energy accumulation and the existing scheme without energy accumulation versus source transmit power $P_s$ for different transmission rate where the interferers' power $P_I = 35dBm$. \label{fig:system}}
\end{figure}
In Fig. 2, we depict the system throughput of the proposed ATF scheme with energy accumulation and the existing one without energy accumulation versus $P_s$ for different transmission rate. The existing scheme in the simulation uses a three-time-slot strategy, where the first time slot is used for energy harvesting, the second and last time slot are used for information transmitting and forwarding by exhausting the harvested energy during the first time slot. It can be observed that implementing energy accumulation at the WEH relay introduces a significant performance gain to the system. Moreover, the gain is increased as the transmission rate grows. This is understandable since energy accumulation can helpfully accumulate the harvested energy and allow the relay to forward information appropriately. The relay can thus schedule the harvested energy more properly and avoid the high energy consumption required by a higher transmission rate in bad channel conditions.



\section{Conclusions}
In this paper, we proposed an accumulate-then-forward scheme for a wireless energy harvesting (WEH) MIMO relay network suffering from co-channel interference. We modeled the dynamic charging and discharging behaviors of the finite-capacity relay battery by a a finite-state Markov chain (MC). We evaluated the transition probability matrix of the MC and derived a closed-form expression of system throughput over mixed Nakagami-m and Rayleigh fading channels. Numerical results validated the theoretical analysis and demonstrated the impact of co-channel interference on the system performance. Specifically, different from the conventional communication networks where CCI is always harmful, the CCI may benefit the performance of a WEH relay network. Results also showed that the proposed ATF scheme with energy accumulation outperforms the existing one without energy accumulation significantly.


\ifCLASSOPTIONcaptionsoff
  \newpage
\fi


\bibliographystyle{IEEEtran}
\bibliography{References}

%

\end{document}